\pdfoutput=1

\documentclass[12pt,a4paper]{article}

\usepackage{ifthen} 
\newboolean{pdflatex}
\setboolean{pdflatex}{true} 

\newboolean{articletitles}
\setboolean{articletitles}{true} 

\newboolean{uprightparticles}
\setboolean{uprightparticles}{false} 

\usepackage{microtype}
\usepackage{lineno}  
\usepackage{xspace} 
\usepackage{afterpage}                    

\usepackage{graphicx}  
\usepackage{color}
\usepackage{colortbl}
\usepackage{pinlabel}
\usepackage{multirow}

\usepackage{caption} 

\usepackage{amsmath} 
\usepackage{amssymb}
\usepackage{amsfonts}
\usepackage{upgreek} 

\newcommand*\patchAmsMathEnvironmentForLineno[1]{%
\expandafter\let\csname old#1\expandafter\endcsname\csname #1\endcsname
\expandafter\let\csname oldend#1\expandafter\endcsname\csname
end#1\endcsname
 \renewenvironment{#1}%
   {\linenomath\csname old#1\endcsname}%
   {\csname oldend#1\endcsname\endlinenomath}%
}
\newcommand*\patchBothAmsMathEnvironmentsForLineno[1]{%
  \patchAmsMathEnvironmentForLineno{#1}%
  \patchAmsMathEnvironmentForLineno{#1*}%
}
\AtBeginDocument{%
\patchBothAmsMathEnvironmentsForLineno{equation}%
\patchBothAmsMathEnvironmentsForLineno{align}%
\patchBothAmsMathEnvironmentsForLineno{flalign}%
\patchBothAmsMathEnvironmentsForLineno{alignat}%
\patchBothAmsMathEnvironmentsForLineno{gather}%
\patchBothAmsMathEnvironmentsForLineno{multline}%
}

\usepackage{hyperref}    
\usepackage[all]{hypcap} 

\def\lhcb {\mbox{LHCb}\xspace}

\ifthenelse{\boolean{uprightparticles}}%
{

\def\Pmu         {\ensuremath{\upmu}\xspace}

\def\Ppi         {\ensuremath{\uppi}\xspace}                
                
\def\Prho        {\ensuremath{\uprho}\xspace}

\def\Pphi        {\ensuremath{\upphi}\xspace}

\def\Pomega      {\ensuremath{\upomega}\xspace}                
 
\def\PDelta      {\ensuremath{\Delta}\xspace}                
\def\PXi      {\ensuremath{\Xi}\xspace}                
\def\PLambda      {\ensuremath{\Lambda}\xspace}                
\def\PSigma      {\ensuremath{\Sigma}\xspace}                
\def\POmega      {\ensuremath{\Omega}\xspace}                
\def\PUpsilon      {\ensuremath{\Upsilon}\xspace}                
 
\def\PB      {\ensuremath{\mathrm{B}}\xspace}                
                
\def\PD      {\ensuremath{\mathrm{D}}\xspace}

\def\PK      {\ensuremath{\mathrm{K}}\xspace}

\def\Pb      {\ensuremath{\mathrm{b}}\xspace}                
\def\Pc      {\ensuremath{\mathrm{c}}\xspace}

\def\Pi      {\ensuremath{\mathrm{i}}\xspace}

\def\Pu      {\ensuremath{\mathrm{u}}\xspace}

}
{

\def\Pmu         {\ensuremath{\mu}\xspace}

\def\Ppi         {\ensuremath{\pi}\xspace}                
                
\def\Prho        {\ensuremath{\rho}\xspace}

\def\Pphi        {\ensuremath{\phi}\xspace}

\def\Pomega      {\ensuremath{\omega}\xspace}                
\mathchardef\PDelta="7101
\mathchardef\PXi="7104
\mathchardef\PLambda="7103
\mathchardef\PSigma="7106
\mathchardef\POmega="710A
\mathchardef\PUpsilon="7107
                
\def\PB      {\ensuremath{B}\xspace}                
                
\def\PD      {\ensuremath{D}\xspace}

\def\PK      {\ensuremath{K}\xspace}

\def\Pb      {\ensuremath{b}\xspace}                
\def\Pc      {\ensuremath{c}\xspace}

\def\Pi      {\ensuremath{i}\xspace}

\def\Pu      {\ensuremath{u}\xspace}

}
 
\def\mmu        {\ensuremath{\Pmu}\xspace}
\def\mup        {\ensuremath{\Pmu^+}\xspace}
\def\mumu       {\ensuremath{\Pmu^+\Pmu^-}\xspace}

\def\uquark    {\ensuremath{\Pu}\xspace}

\def\cquark    {\ensuremath{\Pc}\xspace}

\def\bquark    {\ensuremath{\Pb}\xspace}

\def\pion  {\ensuremath{\Ppi}\xspace}

\def\pip   {\ensuremath{\pion^+}\xspace}
\def\pim   {\ensuremath{\pion^-}\xspace}

\def\kaon  {\ensuremath{K}\xspace}
  \def\Kbar  {\kern 0.2em\overline{\kern -0.2em \PK}{}\xspace}

\def\Kp    {\ensuremath{\kaon^+}\xspace}
\def\Km    {\ensuremath{\kaon^-}\xspace}

  \def\Dbar    {\kern 0.2em\overline{\kern -0.2em \PD}{}\xspace}
\def\D       {\ensuremath{D}\xspace}

\def\Dz      {\ensuremath{\D^0}\xspace}

\def\Dstarp  {\ensuremath{\D^{*+}}\xspace}

\def\B       {\ensuremath{B}\xspace}
\def\Bbar    {\ensuremath{\kern 0.18em\overline{\kern -0.18em \PB}{}}\xspace}

  \def\Y#1S{\ensuremath{\PUpsilon{(#1S)}}\xspace}

\def\proton      {\ensuremath{p}\xspace}

\def\Lz {\ensuremath{\mathit{\Lambda}}\xspace}
\def\Lbar {\ensuremath{\kern 0.1em\overline{\kern -0.1em\PLambda}}\xspace}

\def\Lc      {\ensuremath{\Lz^+_\cquark}\xspace}

\def\BF         {{\ensuremath{\cal B}\xspace}}

\newcommand{\decay}[2]{\ensuremath{#1\!\to #2}\xspace}         

\def\to                 {\ensuremath{\rightarrow}\xspace}
 
\def\AT#1     {\ensuremath{A_{\mathrm{T}}^{#1}}\xspace}           

\def\C#1      {\ensuremath{\mathcal{C}_{#1}}\xspace}                       
\def\Cp#1     {\ensuremath{\mathcal{C}_{#1}^{'}}\xspace}                    
\def\Ceff#1   {\ensuremath{\mathcal{C}_{#1}^{\mathrm{(eff)}}}\xspace}        
\def\Cpeff#1  {\ensuremath{\mathcal{C}_{#1}^{'\mathrm{(eff)}}}\xspace}       
\def\Ope#1    {\ensuremath{\mathcal{O}_{#1}}\xspace}                       
\def\Opep#1   {\ensuremath{\mathcal{O}_{#1}^{'}}\xspace}                    
 
\newcommand{\tev}{\ensuremath{\mathrm{\,Te\kern -0.1em V}}\xspace}
\newcommand{\gev}{\ensuremath{\mathrm{\,Ge\kern -0.1em V}}\xspace}
\newcommand{\mev}{\ensuremath{\mathrm{\,Me\kern -0.1em V}}\xspace}
\newcommand{\kev}{\ensuremath{\mathrm{\,ke\kern -0.1em V}}\xspace}
\newcommand{\ev}{\ensuremath{\mathrm{\,e\kern -0.1em V}}\xspace}
\newcommand{\gevc}{\ensuremath{{\mathrm{\,Ge\kern -0.1em V\!/}c}}\xspace}
\newcommand{\mevc}{\ensuremath{{\mathrm{\,Me\kern -0.1em V\!/}c}}\xspace}
\newcommand{\gevcc}{\ensuremath{{\mathrm{\,Ge\kern -0.1em V\!/}c^2}}\xspace}
\newcommand{\gevgevcccc}{\ensuremath{{\mathrm{\,Ge\kern -0.1em V^2\!/}c^4}}\xspace}
\newcommand{\mevcc}{\ensuremath{{\mathrm{\,Me\kern -0.1em V\!/}c^2}}\xspace}
 
\def\mum  {\ensuremath{{\,\upmu\rm m}}\xspace}

\def\invfb   {\ensuremath{\mbox{\,fb}^{-1}}\xspace}
 
\newcommand{\chisq}{\ensuremath{\chi^2}\xspace}

\newcommand{\chisqip}{\ensuremath{\chi^2_{\rm IP}}\xspace}

\def\gsim{{~\raise.15em\hbox{$>$}\kern-.85em
         \lower.35em\hbox{$\sim$}~}\xspace}
\def\lsim{{~\raise.15em\hbox{$<$}\kern-.85em
         \lower.35em\hbox{$\sim$}~}\xspace}
 
\def\ptot       {\mbox{$p$}\xspace}
\def\pt         {\mbox{$p_{\rm T}$}\xspace}

\def\dllkpi     {\ensuremath{\mathrm{DLL}_{\kaon\pion}}\xspace}
\def\dllppi     {\ensuremath{\mathrm{DLL}_{\proton\pion}}\xspace}

\def\dllmupi    {\ensuremath{\mathrm{DLL}_{\mmu\pi}}\xspace}
 
\def\evtgen     {\mbox{\textsc{EvtGen}}\xspace}

\def\gauss      {\mbox{\textsc{Gauss}}\xspace}
\def\geant      {\mbox{\textsc{Geant4}}\xspace}

\def\pythia     {\mbox{\textsc{Pythia}}\xspace}

\def\tell1  {TELL1\xspace}
\def\ukl1   {UKL1\xspace}

\def\Dppmm  {\ensuremath{\decay{\Dz}{\pip\pim\mumu}}\xspace}			
\def\Dppmmnorm  {\ensuremath{\decay{\Dz}{\pip\pim\phi(\to\mumu)}}\xspace}			
\def\Drhoppmm  {\ensuremath{\decay{\Dz}{\rho(\to\pip\pim)\mumu}}\xspace}			
\def\Dppmmnormhad  {\ensuremath{\decay{\Dz}{\pip\pim\phi(\to\Kp\Km)}}\xspace}			
\def\Dpppp  {\ensuremath{\decay{\Dz}{\pip\pim\pip\pim}}\xspace}			

\def\Mmumu   {\ensuremath{m(\mumu)}\xspace}					

\def\cls     {\ensuremath{\mathrm{CL}_{s}}\xspace}					 
\def\clb     {\ensuremath{\mathrm{CL}_{b}}\xspace}					 
\def\clsb     {\ensuremath{\mathrm{CL_{s+b}}}\xspace}				 
\def\cl     {\ensuremath{\mathrm{CL}}\xspace}						 

\usepackage{tikz}
\usetikzlibrary{decorations.pathreplacing,decorations.pathmorphing,decorations.markings,trees,calc,patterns}
\usepackage{hepnames}
\usepackage{cleveref}
\usepackage{cite} 
\usepackage{mciteplus}
\mciteErrorOnUnknownfalse

\begin{document}

\renewcommand{\thefootnote}{\fnsymbol{footnote}}
\setcounter{footnote}{1}


\begin{titlepage}
\pagenumbering{roman}

\vspace*{-1.5cm}
\centerline{\large EUROPEAN ORGANIZATION FOR NUCLEAR RESEARCH (CERN)}
\vspace*{1.5cm}
\hspace*{-0.5cm}
\begin{tabular*}{\linewidth}{lc@{\extracolsep{\fill}}r}
\ifthenelse{\boolean{pdflatex}}
{\vspace*{-2.7cm}\mbox{\!\!\!\includegraphics[width=.14\textwidth]{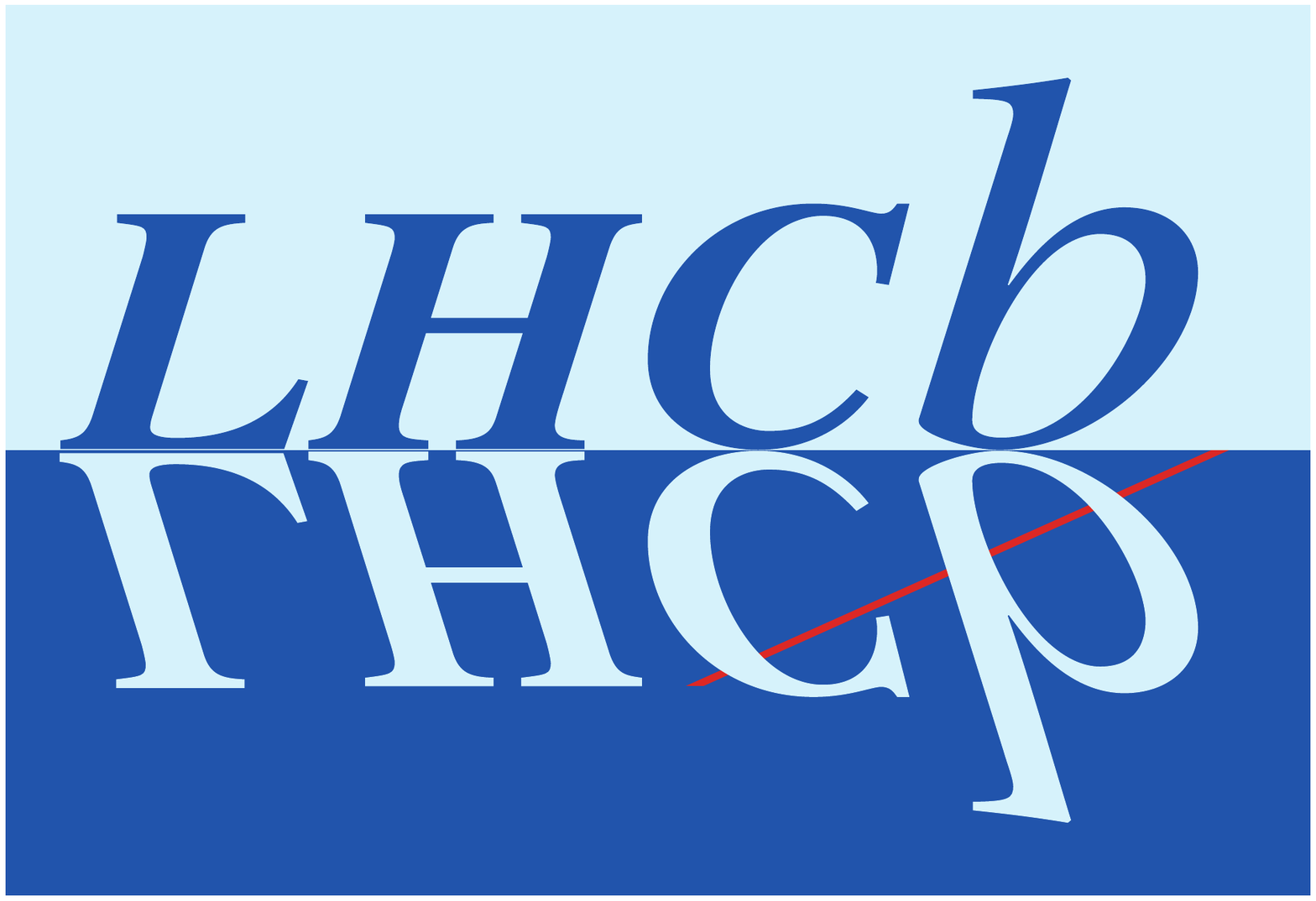}} & &}%
{\vspace*{-1.2cm}\mbox{\!\!\!\includegraphics[width=.12\textwidth]{lhcb-logo.eps}} & &}%
\\
 & & CERN-PH-EP-2013-183 \\  
 & & LHCb-PAPER-2013-050 \\  
 & & 9 October 2013 \\ 
 & & \\
\end{tabular*}

\vspace*{2.0cm}

{\bf\boldmath\huge
\begin{center}
  Search for the decay \Dppmm
\end{center}
}

\vspace*{1.0cm}

\begin{center}
The LHCb collaboration\footnote{Authors are listed on the following pages.}
\end{center}

\vspace{\fill}

\begin{abstract}
\noindent
A search for the \Dppmm decay, where the muon pair does not originate from a resonance, is performed using proton-proton collision data corresponding to an integrated luminosity of $1.0\invfb$ recorded by the LHCb experiment at a centre-of-mass energy of $7\tev$.
No signal is observed and an upper limit on the relative branching fraction with respect to the resonant decay mode \Dppmmnorm, under the assumption of a phase-space model, is found to be
\begin{equation*}
\mathcal{\BF}(\Dppmm)/\mathcal{\BF}(\Dz\to\pip\pim \phi (\to \mup \mu^-)) < 0.96\\
\end{equation*}
at $90\%$ confidence level. The upper limit on the absolute branching fraction is evaluated to be $\mathcal{\BF}(\Dppmm) < 5.5 \, \times 10^{-7}$ at 90\% confidence level.
This is the most stringent to date.
\end{abstract}
\vspace*{1.0cm}

\begin{center}
Submitted to Phys. Lett. B
\end{center}

\vspace{\fill}

{\footnotesize 
\centerline{\copyright~CERN on behalf of the \lhcb collaboration, license \href{http://creativecommons.org/licenses/by/3.0/}{CC-BY-3.0}.}}
\vspace*{2mm}

\end{titlepage}


\newpage
\setcounter{page}{2}
\mbox{~}
\newpage
%
\centerline{\large\bf LHCb collaboration}
\begin{flushleft}
\small
R.~Aaij$^{40}$, 
B.~Adeva$^{36}$, 
M.~Adinolfi$^{45}$, 
C.~Adrover$^{6}$, 
A.~Affolder$^{51}$, 
Z.~Ajaltouni$^{5}$, 
J.~Albrecht$^{9}$, 
F.~Alessio$^{37}$, 
M.~Alexander$^{50}$, 
S.~Ali$^{40}$
% ...

{\footnotesize \it
% ...
$ ^{5}$Clermont Universit\'{e}, Universit\'{e} Blaise Pascal, CNRS/IN2P3, LPC, Clermont-Ferrand, France\\
$ ^{6}$CPPM, Aix-Marseille Universit\'{e}, CNRS/IN2P3, Marseille, France\\
% ...
$ ^{9}$Fakult\"{a}t Physik, Technische Universit\"{a}t Dortmund, Dortmund, Germany\\
% ...
$ ^{36}$Universidad de Santiago de Compostela, Santiago de Compostela, Spain\\
$ ^{37}$European Organization for Nuclear Research (CERN), Geneva, Switzerland\\
% ...
$ ^{40}$Nikhef National Institute for Subatomic Physics, Amsterdam, The Netherlands\\
% ...
$ ^{45}$H.H. Wills Physics Laboratory, University of Bristol, Bristol, United Kingdom\\
% ...
$ ^{50}$School of Physics and Astronomy, University of Glasgow, Glasgow, United Kingdom\\
$ ^{51}$Oliver Lodge Laboratory, University of Liverpool, Liverpool, United Kingdom\\
% ...
}
\end{flushleft}

%
\cleardoublepage


\renewcommand{\thefootnote}{\arabic{footnote}}
\setcounter{footnote}{0}



\pagestyle{plain} 
\setcounter{page}{1}
\pagenumbering{arabic}

\linenumbers

%

\tikzset{
photon/.style={decorate, decoration={snake}, draw=red},
particle/.style={draw=blue, postaction={decorate},decoration={markings,mark=at position .5 with {\arrow[draw=blue]{>}}}},
antiparticle/.style={draw=blue, postaction={decorate},decoration={markings,mark=at position .5 with {\arrow[draw=blue]{<}}}}, 
gluon/.style={decorate, draw=black,decoration={snake,amplitude=4pt, segment length=5pt}}, 
majorana/.style={draw=black, postaction={decorate},decoration={markings,mark=at position .48 with {\arrow[draw=black]{>}},mark=at position .52 with {\arrow[draw=black]{<}}}},
gluonloop/.style={circle, decorate, draw=black, decoration={coil,aspect=1.2,amplitude=2pt, segment length=4pt},minimum height=1.2em},
}

\section{Introduction}
\label{sec:Introduction}

Flavour-changing neutral current (FCNC) processes are rare within the Standard Model (SM) as they cannot occur at tree level and are suppressed by the Glashow-Iliopoulos-Maiani (GIM) mechanism at loop level. In contrast to the \B meson system, where the high mass of the top quark in the loop weakens the suppression, the GIM cancellation is almost exact~\cite{Fajfer:2006yc} in \D meson decays, leading to expected branching fractions for $\cquark\to\uquark\mumu$ processes in the range $(1-3)\times10^{-9}$~\cite{Fajfer:2007dy,Paul:2011ar,Cappiello:2012vg}. This suppression allows for sub-leading processes with potential for physics beyond the SM, such as FCNC decays of \D mesons, and the coupling of up-type quarks in electroweak processes illustrated in Fig.~\ref{fig:FDs}, to be probed more precisely.

The total branching fraction for these decays is expected to be dominated by long-distance contributions involving resonances, such as $\D^0 \to \pip \pim V( \to \mup \mu^- )$, where $V$ can be any of the light vector mesons $\phi$, $\rho^0$ or $\omega$. The corresponding branching fractions can reach ${\cal{O}}(10^{-6})$~\cite{Fajfer:2007dy,Paul:2011ar,Cappiello:2012vg}. The angular structure of these four-body semileptonic \Dz decays provides access to a variety of differential distributions. Of particular interest are angular asymmetries that allow for a theoretically robust separation of long- and short-distance effects, the latter being more sensitive to physics beyond the SM \cite{Cappiello:2012vg}.
No such decays have been observed to date and the most stringent limit reported is $\BF(\Dppmm) < 3.0 \times 10^{-5}$ at 90\% confidence level (\cl) by the E791 collaboration~\cite{Aitala:2000kk}.
The same processes can be probed using $D^+_{(s)} \to \pi^+ \mu^+\mu^-$ decays. Upper limits on their branching fractions have been recently set to $\BF(D^+ \to \pi^+ \mu^+\mu^-) < 7.3\times 10^{-8}$ and $\BF(D^+_{s} \to \pi^+ \mu^+\mu^-) < 4.1\times 10^{-7}$ at 90\% CL by the LHCb collaboration \cite{d2pimumu}.

This Letter presents the result of a search for the \Dppmm decay, in which the muons do not originate from a resonance, performed using $\Dstarp \to \Dz \pip$ decays, with the \Dstarp meson produced  directly at the $pp$ collision primary vertex. The reduction in background yield associated with this selection vastly compensates for the loss of signal yield. No attempt is made to distinguish contributions from intermediate resonances in the dipion invariant mass such as the $\rho^0$.
Throughout this Letter, the inclusion of charge conjugate processes is implied.
The data samples used in this analysis correspond to an integrated luminosity of 1.0\invfb at $\sqrt{s}=7$\tev recorded by the LHCb experiment.
\begin{figure}[!hb]
\centering

\begin{tikzpicture}[scale=1.0]
\begin{scope}
\coordinate (a) at (0,1); 
\coordinate (b) at (0,0); 
\coordinate (c) at (4,1); 
\coordinate (d) at (4,-2); 
\coordinate (e) at (1.2,1); 
\coordinate (f) at (2,1); 
\coordinate (g) at (2.7,1); 
\coordinate (h) at (3,2.5); 
\coordinate (i) at (4,2); 
\coordinate (j) at (4,3); 
\coordinate (k) at (4,-1); 
\coordinate (l) at (4,0); 
\coordinate (m) at (3.5,-0.5); 
\draw[particle] (a) -- (f);
\draw[particle] (f) -- (c);
\draw[antiparticle] (b)  to [out=0,in=160]  (d);
\draw[photon] (e) to [out=270,in=270]  (g) ; 
\draw[photon] (f) -- (h);
\draw[particle] (h) -- (i);
\draw[antiparticle] (h) -- (j);
\draw[antiparticle] (k) to [out=180,in=270] (m) to [out=90,in=180] (l);
\node at ($(a)$) [label={[label distance=-1.5mm] left:$\Pqc$}] {};
\node at ($(b)$) [label={[label distance=-1.5mm] left:$\APqu$}] {};
\node at ($(c)$) [label={[label distance=-1.5mm] right:$\Pqu$}] {};
\node at ($(d)$) [label={[label distance=-1.5mm] right:$\APqu$}] {};
\node at ($(f)$) [label={[label distance=5mm] above:$\Pphoton/\PZz$}] {};
\node at ($(f)$) [label={[label distance=4mm] below:$\PWp$}] {};	
\node at ($(i)$) [label={[label distance=-1.5mm] right:$\Pmuon$}] {};
\node at ($(j)$) [label={[label distance=-1.5mm] right:$\APmuon$}] {};

\node at ($(l)$) [label={[label distance=-7.0mm] left:$\APqd$}] {};
\node at ($(k)$) [label={[label distance=-1.5mm] right:$\Pqd$}] {};
\draw [black,decorate,decoration={brace,amplitude=5pt},xshift=-17pt,yshift=0pt]
  (0,0)  -- (0,1) node [black,midway,left=0pt,xshift=-5pt] {$\Dz$};
\draw [black,decorate,decoration={brace,amplitude=5pt},xshift=17pt,yshift=0pt]
  (4,1)  -- (4,0) node [black,midway,right=0pt,xshift=5pt] {$\pip$};

\draw [black,decorate,decoration={brace,amplitude=5pt},xshift=17pt,yshift=0pt]
(4,-1)  -- (4,-2) node [black,midway,right=0pt,xshift=5pt] {$\pim$};
\end{scope}

\begin{scope}[xshift=8 cm]
\coordinate (a) at (0,1); 
\coordinate (b) at (0,0); 
\coordinate (c) at (4,1); 
\coordinate (d) at (4,-2); 
\coordinate (e) at (1,1); 
\coordinate (f) at (2,2); 
\coordinate (g) at (3,1); 
\coordinate (h) at (4,2); 
\coordinate (i) at (3,3); 
\coordinate (j) at (5,3); 
\coordinate (k) at (4,-1); 
\coordinate (l) at (4,0); 
\coordinate (m) at (3.5,-0.5); 
\draw[particle] (a) -- (c);
\draw[antiparticle] (b) to [out=0,in=160] (d);
\draw[photon] (e) -- (f);
\draw[photon] (g) -- (h);
\draw[antiparticle] (f) -- (i);
\draw[particle] (f) -- (h);
\draw[particle] (h) -- (j);
\draw[antiparticle] (k) to [out=180,in=270] (m) to [out=90,in=180] (l);
\node at ($(a)$) [label={[label distance=-1.5mm] left:$\Pqc$}] {};
\node at ($(b)$) [label={[label distance=-1.5mm] left:$\APqu$}] {};
\node at ($(c)$) [label={[label distance=-1.5mm] right:$\Pqu$}] {};
\node at ($(d)$) [label={[label distance=-1.5mm] right:$\APqu$}] {};
\node at ($(e)$) [label={[label distance=0.5mm] above:$\PWp$}] {};
\node at ($(g)$) [label={[label distance=0.5mm] above:$\PWm$}] {};	
\node at ($(i)$) [label={[label distance=-1.5mm] right:$\APmuon$}] {};
\node at ($(j)$) [label={[label distance=-1.5mm] right:$\Pmuon$}] {};

\node at ($(l)$) [label={[label distance=-7.0mm] left:$\APqd$}] {};
\node at ($(k)$) [label={[label distance=-1.5mm] right:$\Pqd$}] {};
\draw [black,decorate,decoration={brace,amplitude=5pt},xshift=-17pt,yshift=0pt]
  (0,0)  -- (0,1) node [black,midway,left=0pt,xshift=-5pt] {$\Dz$};
\draw [black,decorate,decoration={brace,amplitude=5pt},xshift=17pt,yshift=0pt]
  (4,1)  -- (4,0) node [black,midway,right=0pt,xshift=5pt] {$\pip$};
\draw [black,decorate,decoration={brace,amplitude=5pt},xshift=17pt,yshift=0pt]
(4,-1)  -- (4,-2) node [black,midway,right=0pt,xshift=5pt] {$\pim$};
\end{scope}
\end{tikzpicture}

\caption{\small{Leading Feynman diagrams for the FCNC decay \Dppmm in the SM.}}
\label{fig:FDs}
\end{figure}

The analysis is performed in four dimuon mass ranges to exclude decays dominated by the contributions of resonant dimuon final states. The regions at low and high dimuon masses, away from the $\eta$, $\rho^0$ and $\phi$ resonant regions, are the most sensitive to non-SM physics and are defined as the signal regions. The signal yield is normalised to the yield of resonant \Dppmmnorm decays, isolated in an appropriate dimuon range centred around the $\phi$ pole.

\section{The \lhcb detector and trigger}
\label{sec:detector}

The \lhcb detector~\cite{Alves:2008zz} is a single-arm forward
spectrometer covering the \mbox{pseudorapidity} range $2<\eta <5$,
designed for the study of particles containing \bquark or \cquark
quarks. The detector includes a high-precision tracking system
consisting of a silicon-strip vertex detector surrounding the $pp$
interaction region, a large-area silicon-strip detector located
upstream of a dipole magnet with a bending power of about
$4{\rm\,Tm}$, and three stations of silicon-strip detectors and straw
drift tubes placed downstream.
The combined tracking system provides a momentum measurement with
relative uncertainty that varies from 0.4\% at 5\gevc to 0.6\% at 100\gevc,
and impact parameter resolution of 20\mum for
tracks with large transverse momentum. Different types of charged hadrons are distinguished by information
from two ring-imaging Cherenkov detectors~\cite{LHCb-DP-2012-003}. Photon, electron and
hadron candidates are identified by a calorimeter system consisting of
scintillating-pad and preshower detectors, an electromagnetic
calorimeter and a hadronic calorimeter. Muons are identified by a
system composed of alternating layers of iron and multiwire
proportional chambers~\cite{LHCb-DP-2012-002}.

The trigger~\cite{LHCb-DP-2012-004} consists of a
hardware stage, based on information from the calorimeter and muon
systems, followed by a software stage, which applies a full event
reconstruction.
The hardware trigger selects muons with transverse momentum, \pt, exceeding 1.48\gevc, and dimuons whose product of \pt values exceeds $(1.3\gevc)^2$.
In the software trigger, at least one of the final state muons is required to have momentum larger than 8\gevc, and to have an impact parameter, IP, defined as  the minimum distance of the particle trajectory from the associated primary vertex (PV) in three dimensions, greater than 100\mum. 
Alternatively, a dimuon trigger accepts events with oppositely charged muon candidates having good track quality, \pt exceeding $0.5\gevc$, and momentum exceeding $6\gevc$.
In a second stage of the software trigger, two algorithms select \Dppmm candidates. The first algorithm, used to increase the efficiency in the highest dimuon mass region, requires oppositely charged muons with scalar sum of \pt greater than $1.5\gevc$ and dimuon mass greater than $1\gevcc$. A second algorithm selects events with two oppositely charged muons and two oppositely charged hadrons with no invariant mass requirement on the dimuon.

Simulated events for the signal, using a phase-space model, and the normalisation mode, are used to define selection criteria and to evaluate efficiencies. The $pp$ collisions are generated using \pythia 6.4 \cite{Sjostrand:2006za} with a specific LHCb configuration \cite{LHCb-PROC-2010-056}. Decays of hadronic particles are described by \evtgen \cite{Lange:2001uf}. The interaction of the generated particles with the detector and its response are implemented using the \geant toolkit \cite{Allison:2006ve, *Agostinelli:2002hh} as described in Ref.~\cite{LHCb-PROC-2011-006}.

\section{Candidate selection}
\label{sec:candidateselection}
Candidate $\Dppmm$ decays are required to originate from $\Dstarp \to \Dz \pip$ decays.
The \Dz candidate is formed by combining two pion and two muon candidates where both pairs consist of oppositely charged particles. An additional pion track is combined with the \Dz candidate to build the \Dstarp candidate. The \chisq per degree of freedom of the vertex fit is required to be less than 5 for both the \Dstarp and the \Dz candidates.
The angle between the \Dz momentum vector and the direction from the associated PV to the decay vertex, $\theta_{\Dz}$, is required to be less than $0.8^\circ$.
Each of the four particles forming the \Dz meson must have momentum exceeding 3 \gevc and $\pt$ exceeding 0.4 \gevc. The tracks must be displaced with respect to any PV and have \chisqip larger than 4. Here \chisqip is defined as the difference between the $\chisq$ of the PV fit done with and without the track under consideration.

Further discrimination is achieved using a boosted decision tree (BDT) \cite{Breiman, *Roe, Hocker:2007ht}, which distinguishes between signal and combinatorial background candidates. This multivariate analysis algorithm is trained using simulated \Dppmm signal events and a background sample taken from data mass sidebands around the \Dppmm signal mass region. Only 1\% of the candidates in the sidebands are used in the training. The BDT uses the following variables:
$\theta_{\Dz}$, \chisq of the decay vertex and flight distance of the \Dz candidate, 
\ptot and \pt of the \Dz candidate and of each of the four final state tracks, \chisq of the vertex and \pt of the \Dstarp candidate,
\chisqip of the \Dz candidate and of the final state particles,
the maximum distance of closest approach between all pairs of tracks forming the \Dz and \Dstarp candidates,
and the \pt and \chisqip of the bachelor pion from the \Dstarp candidate.

The BDT discriminant is used to classify each candidate. Assuming a signal branching fraction of $10^{-9}$, an optimisation study is performed to choose the combined BDT and muon particle identification (PID) selection criteria that maximise the expected statistical significance of the signal. This significance is defined as $S/\sqrt{S+B}$, where $S$ and $B$ are the signal and background yields respectively.
The PID information is quantified as the difference in the log-likelihood of the detector response under different particle mass hypotheses (DLL) \cite{LHCb-DP-2012-003,LHCb-DP-2013-001}. The optimisation procedure yields an optimal threshold for the BDT discriminant and a minimum value for $\dllmupi$ (the difference between the muon and pion hypotheses) of 1.5 for both \mmu candidates. In addition, the pion candidate is required to have \dllkpi less than 3.0 and \dllppi less than 2.0, and each muon candidate must not share hits in the muon stations with any other muon candidate. In the 2\% of events in which multiple candidates are reconstructed, the candidate with the smallest \Dz vertex \chisq  is chosen.

The bachelor \pip of the $\Dstarp \to \Dz \pip$ decay is constrained to the PV using a Kalman filter \cite{Hulsbergen:2005pu}. This constraint improves the resolution for the mass difference between the \Dstarp and the \Dz candidates, $\Delta m \equiv m(\pi^+\pi^-\mu^+\mu^-\pi^+)-m(\pi^+\pi^-\mu^+\mu^-)$, by a factor of two, down to $0.3$\mevcc. Candidates are selected with a $\Delta m$ value in the range $140.0-151.4$\mevcc.

Candidates from the kinematically similar decay \Dpppp form an important peaking background due to the possible misidentification of two oppositely charged pions as muons.
A sample of this hadronic background is retained with a selection that is identical to that applied to the signal except that no muon identification is required. These candidates are then reconstructed under the \Dppmm hypothesis and a subsample of the candidates, in which at least one such pion satisfies the muon identification requirements, is used to determine the shape of this peaking background in each region of dimuon mass, \Mmumu. Under the correct mass hypotheses the \Dpppp candidates are also used as a control sample to check differences between data and simulation that may affect the event selection performance. Moreover, they are used to determine the expected signal shape in each \Mmumu region by subdividing the \Dpppp sample in the same regions of $m(\pi^+\pi^-)$.

Another potential source of peaking background is due to $\mathit{\Lambda}_c(2595)^+ \to \mathit{\Sigma}_c(2455)^0\pip$ decays, followed by the $\Sigma_c(2455)^0 \to \Lc \pim$ and then $\Lc \to \proton \Km \pip$ decays, with the two pions in the decay chain misidentified as muons and the proton and the kaon misidentified as pions. Therefore, the \dllkpi and \dllppi requirements are tightened to be less than zero for the low-$\Mmumu$ region, where the baryonic background is concentrated, suppressing this background to a negligible level.

Another potentially large background from the $\Dz \to \pip\pim \eta$ decay, followed by the decay $\eta\to\mup\mu^- \gamma$, does not peak at the \Dz mass since candidates in which the $\Mmumu$ is within $\pm20$\mevcc of the nominal $\eta$ mass are removed from the final fit. The remaining contribution to low values of the $m(\pip\pim\mup\mu^-)$ invariant mass is included in the combinatorial background.

\section{Mass fit}
\label{sec:fit}

The shapes and yields of the signal and background contributions are determined using an unbinned maximum likelihood fit to the two-dimensional $\left[ m(\pi^+\pi^-\mu^+\mu^-\pi^+),\Delta m\right]$ distributions in the ranges $1810-1920$ and $140-151.4$ \mevcc, respectively.
This range is chosen to contain all reconstructed \Dppmm candidates.

The \Dppmm data are split into four regions of \Mmumu: two regions containing the $\rho/\omega$ and $\phi$ resonances and two signal regions, referred to as {{low-$\Mmumu$}} and {{high-$\Mmumu$}}, respectively. The definitions of these regions are provided in Table~\ref{tab:yields}. 

The \Dz mass and $\Delta m$ shapes for \Dppmm candidates are described by a double Crystal Ball function \cite{CB1,*CB2}, which consists of a Gaussian core and independent left and right power-law tails, on either sides of the core.
The parameters of these shapes are determined from the \Dpppp control sample independently for each of the four \Mmumu regions.

The \Dpppp peaking background is also split into the predefined dimuon mass regions and is fitted with a double Crystal Ball function. This provides a well-defined shape for this prominent background, which is included in the fit to the signal sample. The yield of the misidentified component is allowed to vary and fitted in each region of the analysis.
The combinatorial background is described by an exponential function in the \Dz candidate mass, while the shape in $\Delta m$ is described by the empirical function $ f_{\Delta}(\Delta m,a)=1-e^{-(\Delta m-\Delta {m_0})/a}$, where the parameter $\Delta {m_0}$ is fixed to $139.6\,\mevcc$.
The two-dimensional shape used in the fit implicitly assumes that $m(\pi^+\pi^-\mu^+\mu^-\pi^+)$ and $\Delta m$ are not correlated.

All the floating coefficients are allowed to vary independently in each of the \Mmumu regions.
Migration between the regions is found to be negligible from simulation studies. The yield observed in the \Pphi region is used to normalise the yields in the signal regions. 

\begin{figure*}[htp]
\centering
\includegraphics[width=0.45\textwidth]{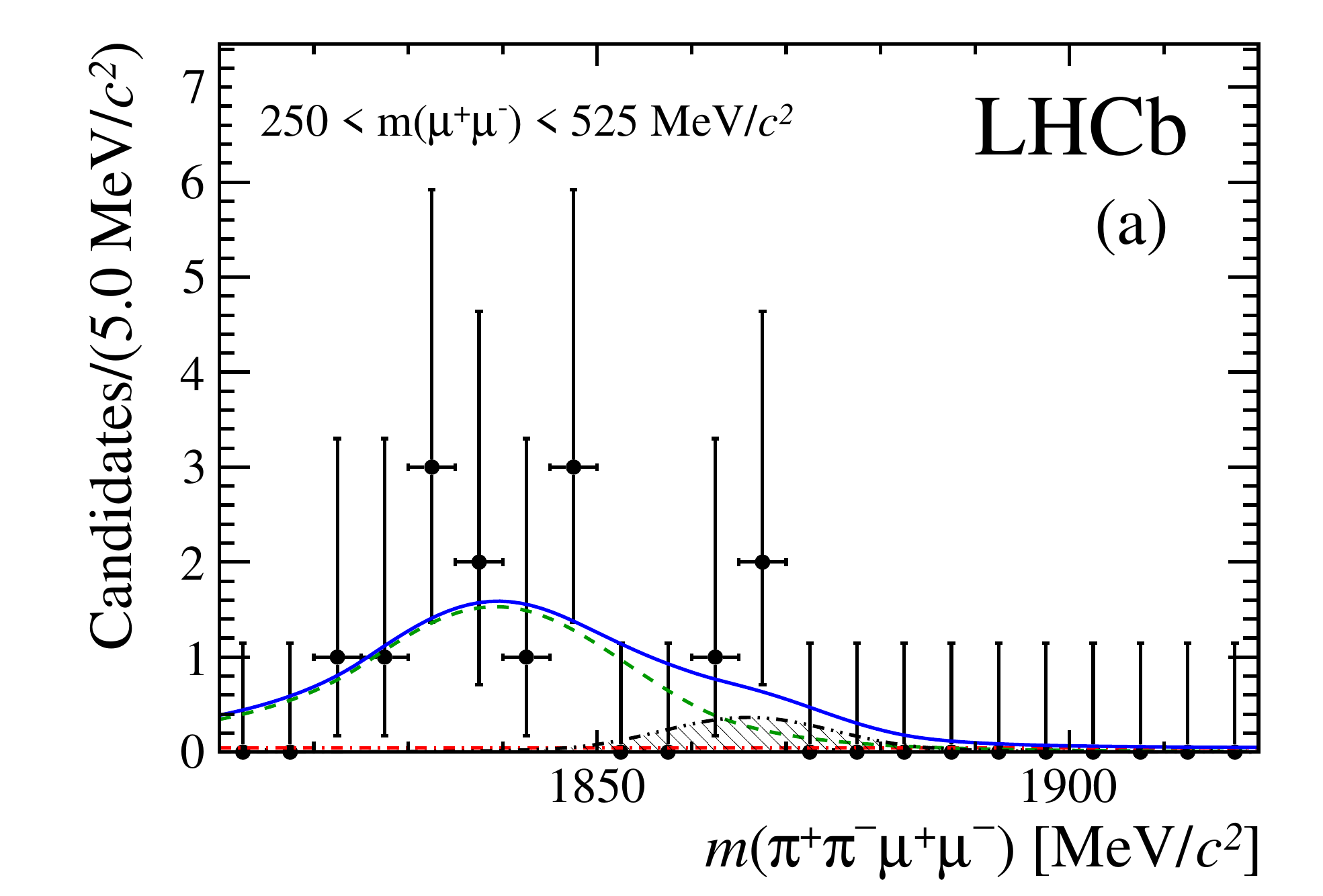}
\includegraphics[width=0.45\textwidth]{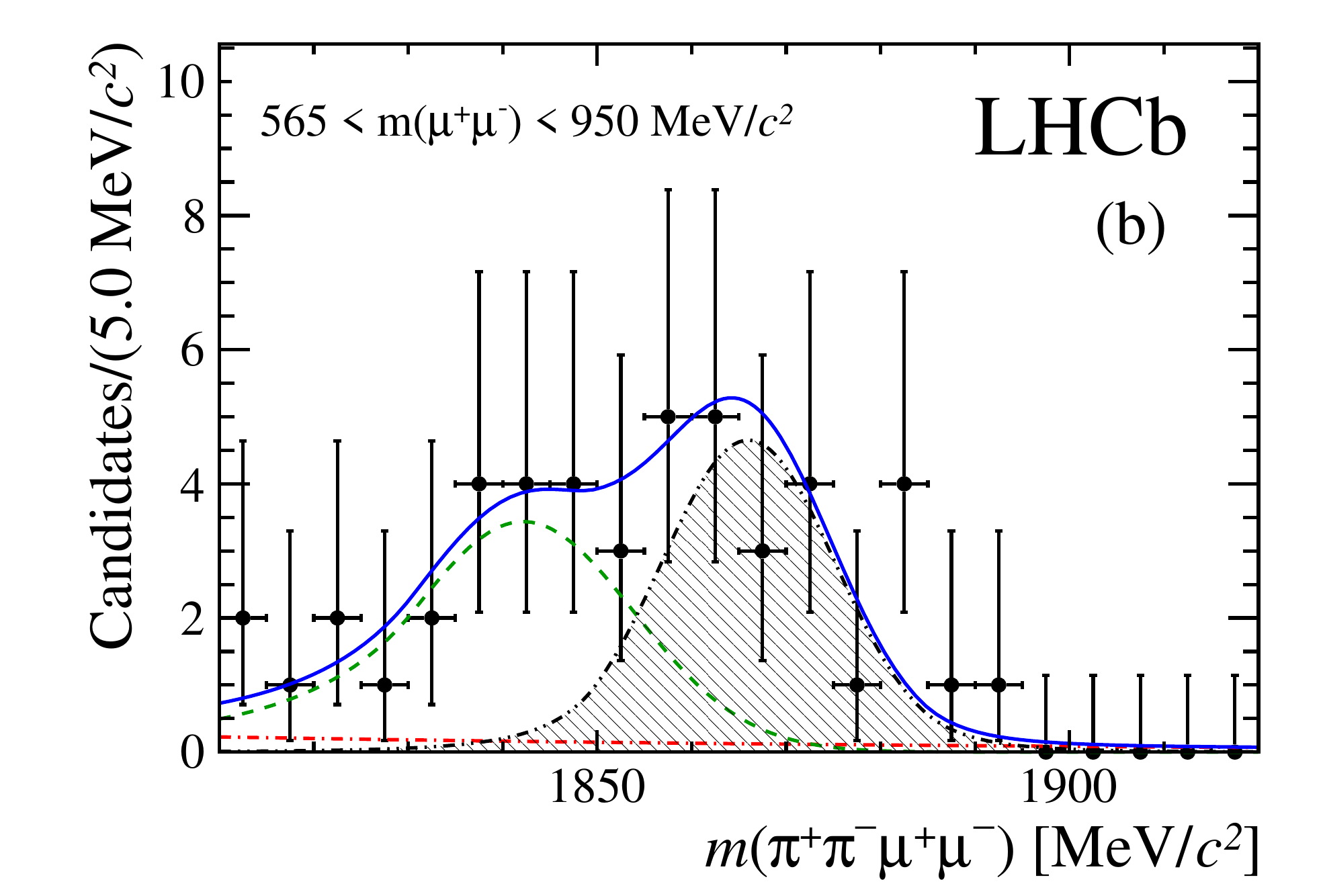}
\includegraphics[width=0.45\textwidth]{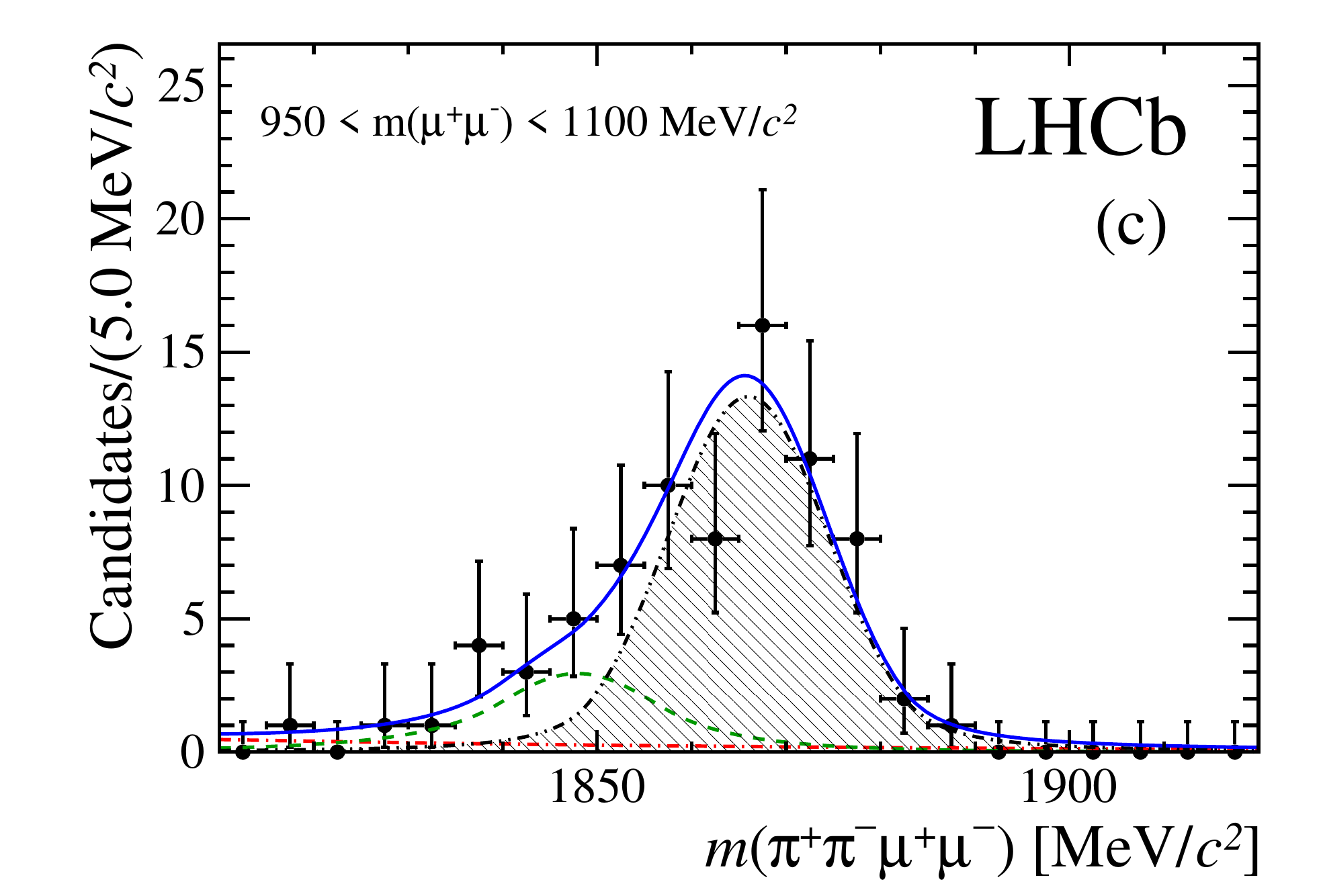}
\includegraphics[width=0.45\textwidth]{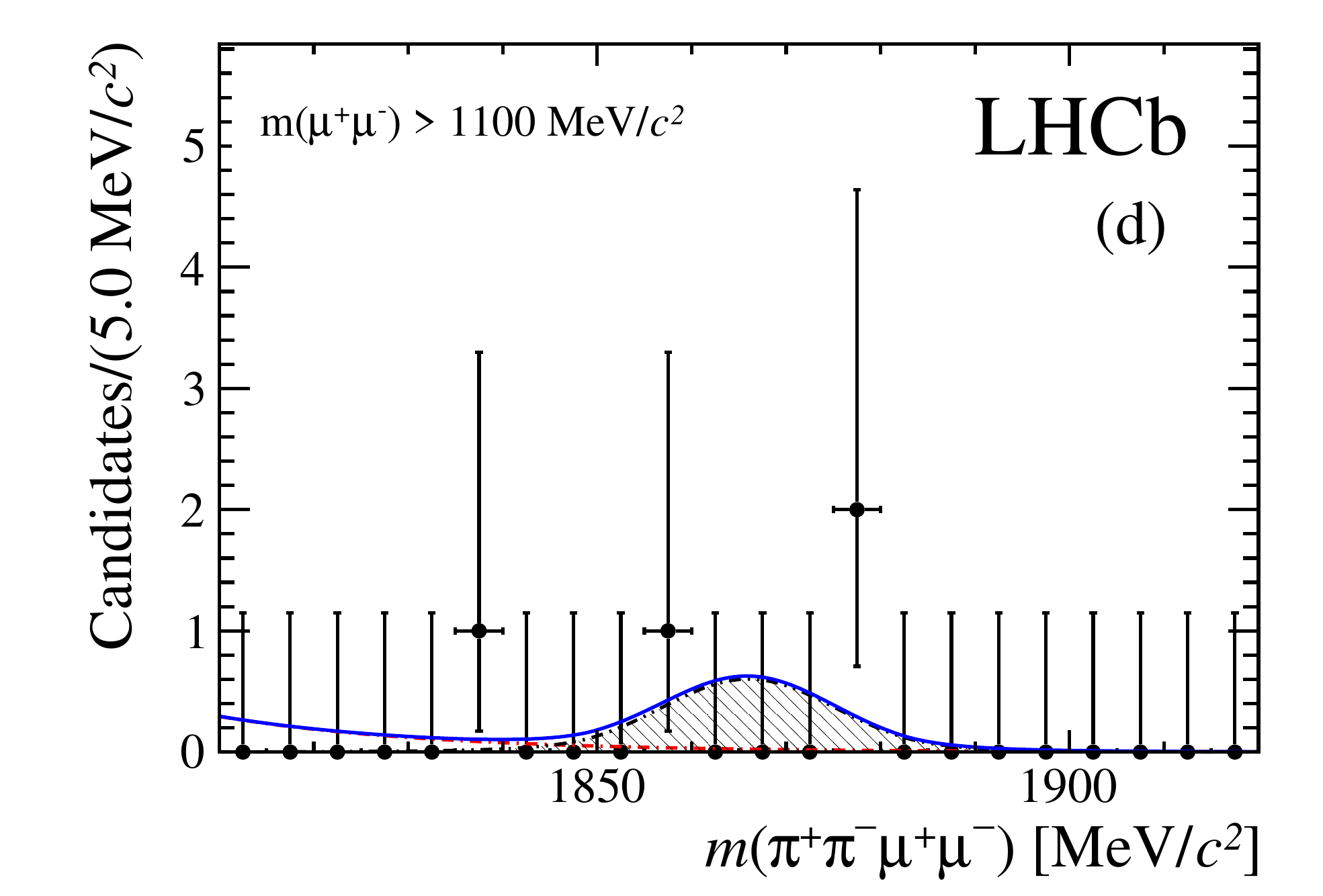}
\caption{\small{Distributions of $m(\pi^+\pi^-\mu^+\mu^-)$ for \Dppmm candidates in the (a) low-\Mmumu, (b) \Prho/\Pomega, (c) \Pphi , and (d) high-\Mmumu regions, with $\Delta m$ in the range $144.4-146.6$ \mevcc. The data are shown as points (black) and the fit result (dark blue line) is overlaid. The components of the fit are also shown: the signal (filled area), the \Dpppp background (green dashed line) and the non-peaking background (red dashed-dotted line).}}
\label{fig:mass1}
\end{figure*}

\begin{figure*}[htp]
\centering
\includegraphics[width=0.45\textwidth]{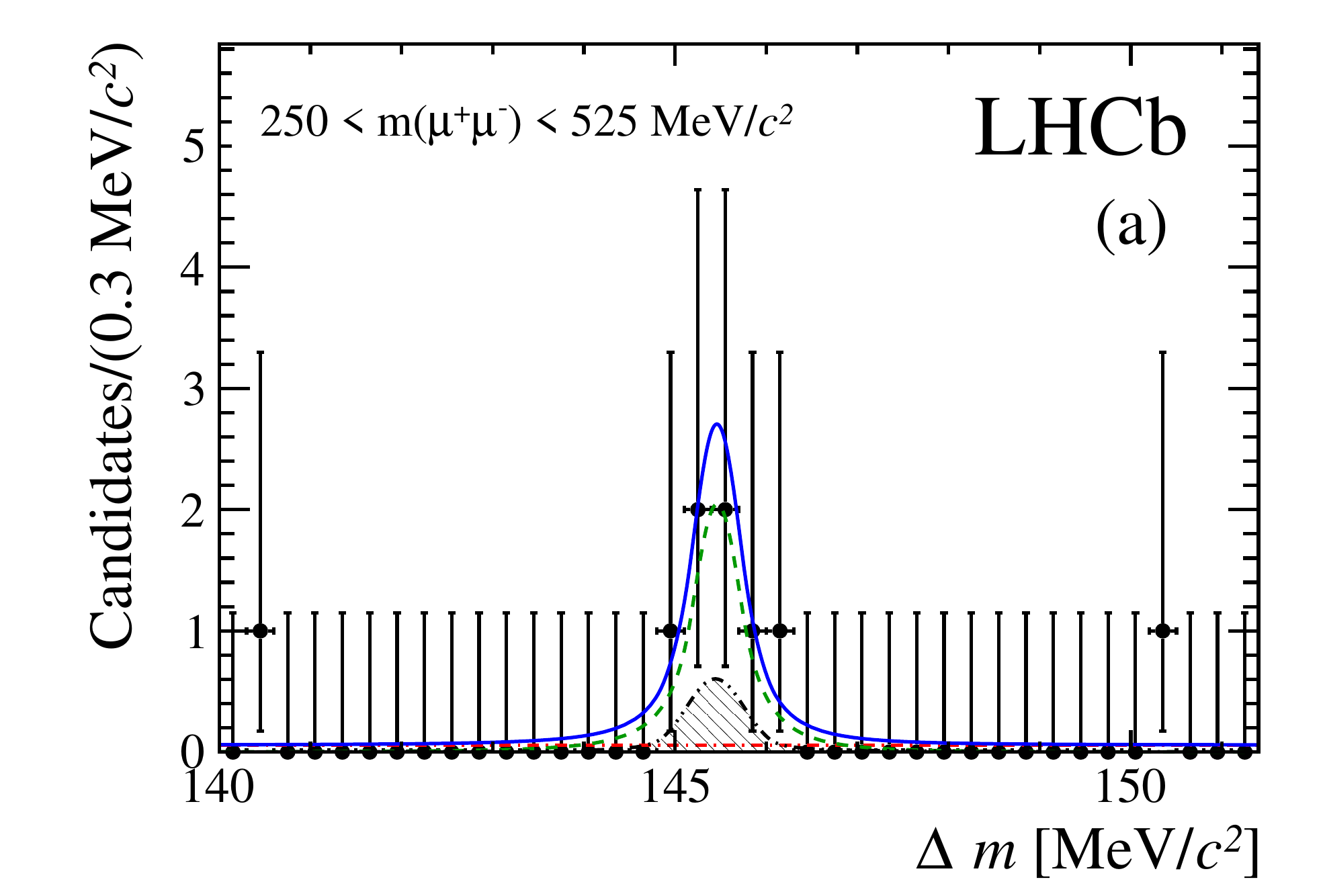}
\includegraphics[width=0.45\textwidth]{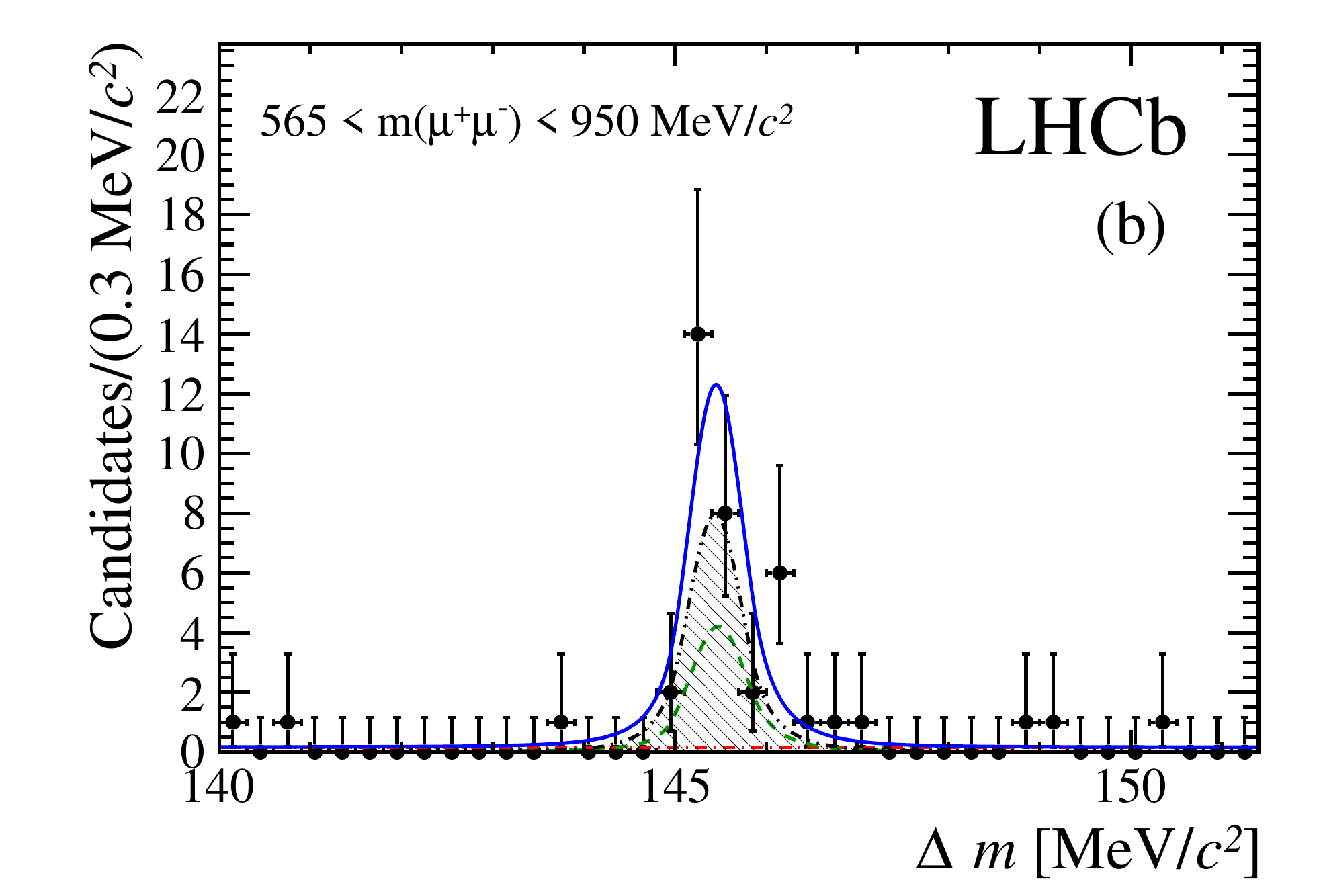}
\includegraphics[width=0.45\textwidth]{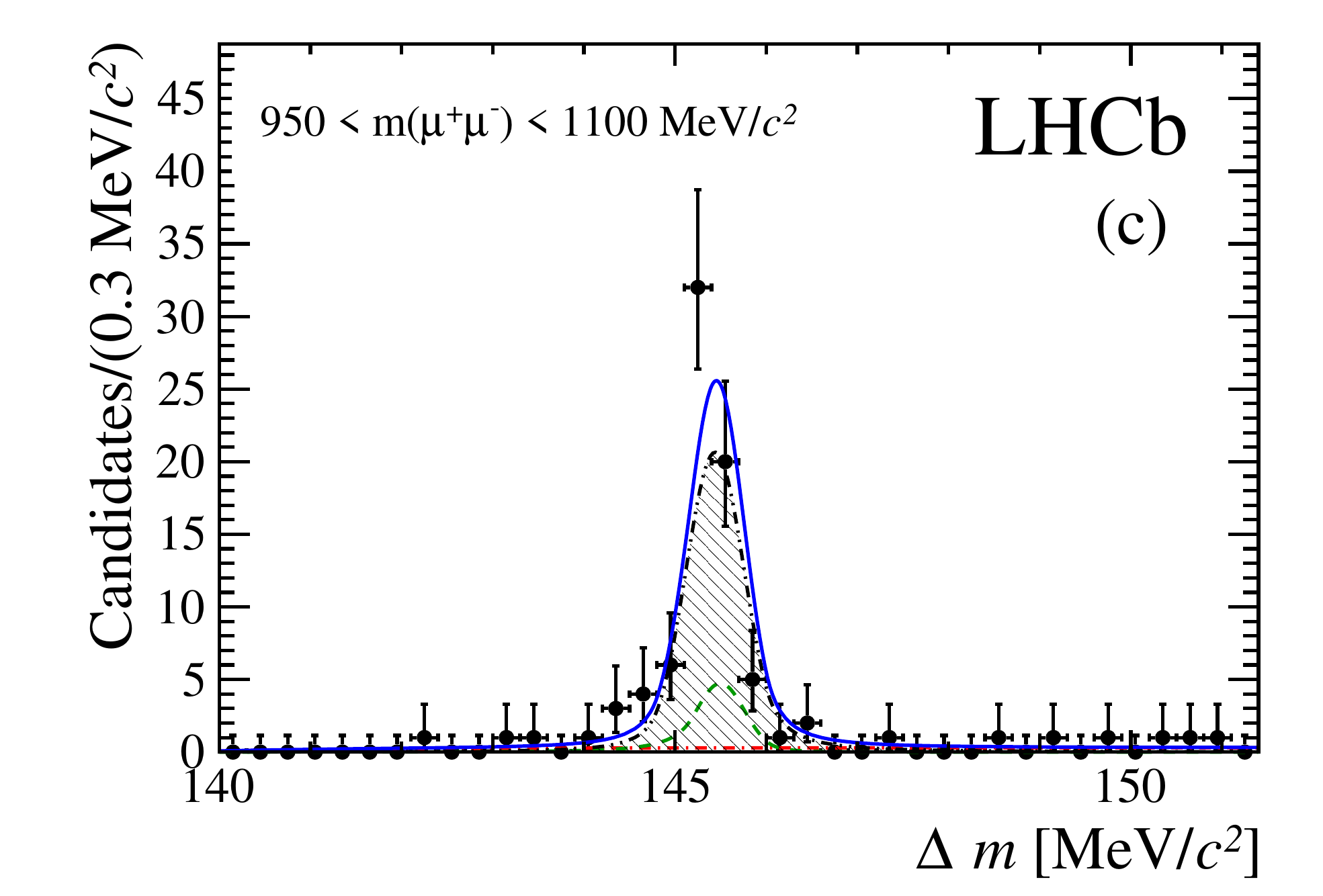}
\includegraphics[width=0.45\textwidth]{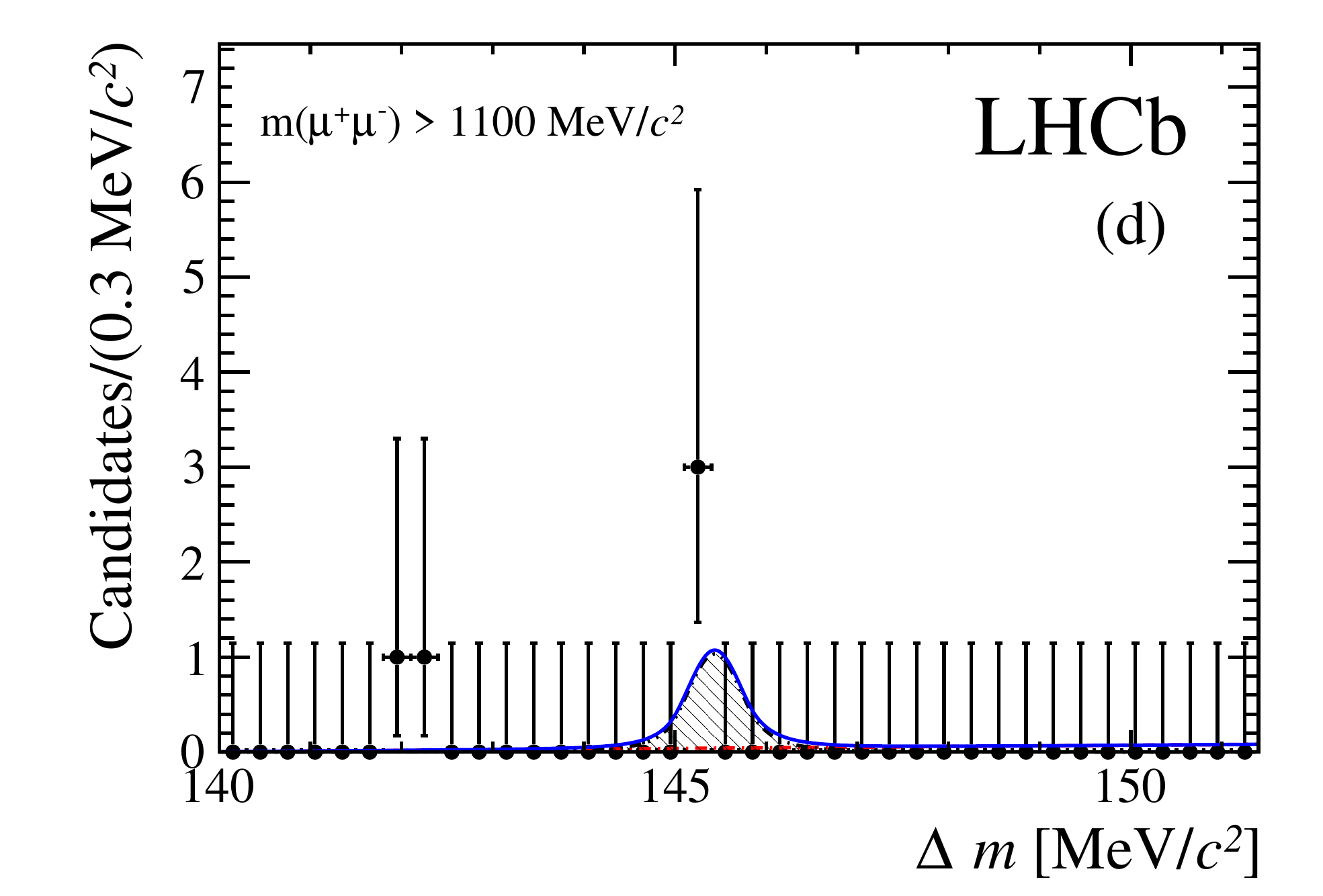}
\caption{\small{Distributions of $\Delta m$ for \Dppmm candidates in the (a) low-\Mmumu, (b) \Prho/\Pomega, (c) \Pphi , and (d) high-\Mmumu regions, with the \Dz invariant mass in the range $1840-1888$ \mevcc. The data are shown as points (black) and the fit result (dark blue line) is overlaid. The components of the fit are also shown: the signal (filled area), the \Dpppp background (green dashed line) and the non-peaking background (red dashed-dotted line).}}
\label{fig:mass2}
\end{figure*}

One-dimensional projections for the \Dz candidate invariant mass and $\Delta m$ spectra, together with the result of the fits, are shown in Figs.~\ref{fig:mass1} and \ref{fig:mass2}, respectively.
The signal yields, which include contributions from the tails of the \Mmumu resonances leaking into the low- and high-\Mmumu ranges, are shown in Table~\ref{tab:yields}. No significant excess of candidates is seen in either of the two signal regions. 

\begin{table}[!t]
\centering
\caption{\small{$D^0\to\pi^+\pi^-\mu^+\mu^-$ fitted yields in the four $m(\mu^+\mu^-)$ regions. The corresponding signal fractions under the assumption of a phase-space model, as described in Section 7, are listed in the last column.}}
\begin{tabular}{lllc}
\hline
Range description \phantom{0} &\phantom{0} \Mmumu [$\mevcc$] \phantom{0}&\phantom{0} \Dppmm yield & \phantom{00}Fraction\\
\hline
\phantom{0} low-\Mmumu & $\phantom{00000-}250-525$ & $\phantom{000000000}2\pm2$ &\phantom{00} 30.6\% \\
\phantom{0} \Prho/\Pomega & $\phantom{00000-}565-950$ & $\phantom{00000000}23\pm6$ &\phantom{00} 43.4\% \\
\phantom{0} \Pphi & $\phantom{00000-}950-1100$ & $\phantom{00000000}63\pm10$ &\phantom{00} 10.1\% \\
\phantom{0} high-\Mmumu & $\phantom{00000-}>1100$ & $\phantom{000000000}3\pm2$ &\phantom{00} 8.9\% \\
\hline
\end{tabular}
\label{tab:yields}
\end{table}

The yields in the signal regions are compatible with the expectations from leakage from the \Mmumu resonant regions. 
The number of expected events from leakage is calculated assuming the \Mmumu spectrum given by a sum of relativistic Breit-Wigner functions, describing the $\eta$, $\rho/\omega$ and $\phi$ resonances. The contribution from each resonance is scaled according to the branching fractions as determined from resonant $\Dz\to\Kp\Km\pip\pim$ and $\Dz\to\pip\pim\pip\pim$ decays \cite{PDG2012}. The resulting shape is used to extrapolate the yields fitted in the $\phi$ and $\rho$ regions into the \Mmumu signal regions. An additional extrapolation is performed using the signal yield in the $\Mmumu$ range $773-793$ \mevcc, where the contribution from the $\omega$ resonance is enhanced. In this approach the interference among different resonances is not accounted for and a systematic uncertainty to the extrapolated yield is assigned according to the spread in their extrapolations. The expected number of leakage events is estimated to be $1\pm1$ in both the low- and high- $\Mmumu$ regions. This precision of this estimate 
is dominated by the systematic uncertainty.

\section{Branching fraction determination}
\label{sec:branchingfractiondetermination}

The \Dppmm branching fraction ratio for each \Mmumu signal region $i$ is calculated using
\begin{equation}
\frac{{\BF(\Dppmm)^i}}{{\BF(\Dppmmnorm)}}
= \frac{N^i_{\Dppmm}}{N_{\Dppmmnorm}}
\times \frac{\epsilon_{\Dppmmnorm}}{\epsilon^i_{\Dppmm}}.
\label{Eq:ExtractBR}
\end{equation}
The yield and efficiency are given by $N_{\Dppmm}$ and $\epsilon_{\Dppmm}$, respectively, for the signal channel, and by $N_{\Dppmmnorm}$ and $\epsilon_{\Dppmmnorm}$ for the reference channel.
The values for the efficiency ratio $\epsilon_{\Dppmm}/\epsilon_{\Dppmmnorm}$ in the low-$\Mmumu$ and high-$\Mmumu$ regions, as estimated from simulations, are $0.24\pm0.03$ and $0.69\pm0.11$, respectively, where the uncertainty reflects the limited statistics of the simulated samples.
The efficiencies for reconstructing the signal decay mode and the reference mode include the geometric acceptance of the detector, the efficiencies for track reconstruction, particle identification, selection and trigger. Both efficiency ratios deviate from unity due to differences in the kinematic distributions of the final state particles in the two decays. Moreover, tighter particle identification requirements are responsible for a lower efficiency ratio in the low-$\Mmumu$ region.
The accuracy with which the simulation reproduces the track reconstruction and particle identification is limited.
Therefore, the corresponding efficiencies are also studied in data and systematic uncertainties are assigned.

An upper limit on the absolute branching fraction is given using an estimate of the branching fraction of the normalisation mode.
The $\Dppmmnorm$ branching fraction is estimated using the results of the amplitude analysis of the $\Dz\to\Kp\Km\pip\pim$ decay performed at CLEO \cite{2012PhRvD..85l2002A}. Only the fit fraction of the decay modes in which the two kaons originate from an intermediate $\phi$ resonance are considered and the $\Dppmmnorm$ branching fraction is calculated by multiplying this fraction by the total $\Dz \to \Kp \Km \pip \pim$ branching fraction and using the known value of $\BF(\phi\to\mup\mu^-)/\BF(\phi\to\Kp\Km)$~\cite{PDG2012}. There are several interfering contributions to the \Dppmmnormhad amplitude. Considering the interference fractions provided in Ref.\cite{2012PhRvD..85l2002A}, the following estimate for the branching fraction is obtained, $\BF(\Dppmmnorm)=(5.2\pm0.6)\times10^{-7}$. This estimate includes only the statistical uncertainty and refers to the baseline fit model used for the CLEO measurement. Similar estimates for $\BF(\Dppmmnorm)$ are performed using all the alternative models 
considered in Ref.\cite{2012PhRvD..85l2002A} assuming the interference fractions to be the same as for the baseline model. The spread among the estimates is used to assign a systematic uncertainty of $17\%$ on $\BF(\Dppmmnorm)$. The above procedure to estimate $\BF(\Dppmmnorm)$ is supported by the narrow width of the \Pphi resonance resulting in interference effects with other channels \cite{2012PhRvD..85l2002A} that are negligible compared to the statistical uncertainty. The estimate for $\BF(\Dppmmnorm)$ is $(5.2\pm1.1)\times10^{-7}$, including both statistical and systematic uncertainties, and is used to set an upper limit on the absolute $\Dppmm$ branching fraction.

A possible alternative normalisation, with respect to the $\rho/\omega$ dimuon mass region, would be heavily limited by the low statistics available and the relatively high contamination from \Dpppp, as can be seen in \figurename{~\ref{fig:mass1}b}.

\section{Systematic uncertainties}
\label{sec:systematics}

Several systematic uncertainties affect the efficiency ratio.
Differences in the particle identification between the signal and the normalisation regions are investigated in data.
A tag-and-probe technique applied to $ b \rightarrow J/\psi X$ decays provides a large sample of muon candidates to determine the muon identification efficiencies \cite{LHCb-DP-2013-001}. General agreement between simulation and data is found to a level of 1\%, which is assigned as a systematic uncertainty.

The particle identification performance for hadrons is investigated by comparing the efficiency in $\Dz\to\pip\pim\pip\pim$ candidates in data and simulation as a function of the \dllkpi requirement. The largest discrepancy between data and simulation on the efficiency ratio is found to be 4\% and is taken as a systematic uncertainty.

Several quantities, particularly the impact parameter, are known to be imperfectly reproduced in the simulation. Since this may affect the reconstruction and selection efficiency, a systematic uncertainty is estimated by smearing track properties to reproduce the distributions observed in data.
The corresponding variation in the efficiency ratio yields an uncertainty of 5\%.  The BDT description in simulation is checked using background-subtracted \Dpppp candidates where no significant difference is seen. Therefore, no extra systematic uncertainty is assigned.

The systematic uncertainty due to possible mismodelling of the trigger efficiency in the simulation is assigned as follows. The trigger requirements in simulations are varied reproducing the typical changes of trigger configurations that occurred during data taking and an alternate efficiency ratio is calculated in both the \Mmumu signal regions. The largest difference between the alternate and the baseline efficiency ratio, 5\%, is found in the low-\Mmumu region. This difference is assumed as the overall systematic uncertainty on the trigger efficiency.

The uncertainties on the efficiency ratio due to the finite size of the simulated samples in the low- and high- \Mmumu regions are 12\% and 16\% respectively. The production of significantly larger sample of simulated events is impractical due to the low reconstruction and selection efficiencies, particularly in the signal regions. In addition, the statistical uncertainties of the fitted yields in data, listed in \tablename{~\ref{tab:yields}}, dominate the total uncertainty.
The sources of uncertainty are summarised in Table~\ref{tab:systematics1}.

\begin{table}[!b]
\centering 
\caption{\small{Relative systematic uncertainties averaged over all the \Mmumu regions for the efficiency ratio.}} 
\begin{tabular}{lc} 
\hline
Source &  Uncertainty (\%)\\
\hline 
Trigger efficiency & 5 \\
Hadron identification & 4 \\
Reconstruction and selection efficiency & 5 \\
Muon identification & 1 \\
Finite simulation sample size& 12\textendash16\\
\hline
Total & 15\textendash18 \\
\hline
\end{tabular}
\label{tab:systematics1} 
\end{table}

According to simulations, biases in the efficiency ratio introduced by varying the relative contribution of $\Dz\to\rho^0(\to\pi\pi)\phi(\to\mu\mu)$ and three-body \Dppmmnorm decays are well within the assigned uncertainty. Varying  the value of $\BF(\Dppmmnorm)$ has a negligible effect on the number of leakage events, and no additional systematic uncertainty is assigned.

The systematic uncertainties affecting the yield ratio are taken into account when the branching fraction limits are calculated.
The shapes of the signal peaks are taken from the \Dpppp samples separately for each \Mmumu region to account for variations of the shape as a function of \Mmumu.
The impact of alternative shapes for the signal and misidentified \Dpppp decays on the fitted yields and the final limit are investigated. The signal and misidentification background shapes in the signal regions are fitted using the shapes obtained in the $\phi$ region, and from \Dpppp events reconstructed as \Dppmm, but without any muon identification requirements. The change in the result is negligible.

The absolute branching fraction limit includes an extra uncertainty of 21\% from the estimate of the branching fraction of the normalisation mode.

\section{Results}
\label{sec:results}
The compatibility of the observed distribution of candidates with a signal plus background or background-only hypothesis is evaluated using the \cls method~\cite{CLsMethod, Junk:1999kv}, which includes the treatment of systematic uncertainties.
Upper limits on the non-resonant \Dppmm to \Dppmmnorm branching fraction ratio and on the absolute \Dppmm branching fraction are determined using the observed distribution of \cls as a function of the branching fraction in each \Mmumu search region. The extrapolation to the full \Mmumu phase space is performed assuming a four-body phase space model for \Dppmm for which fractions in each \Mmumu region are quoted in \tablename{~\ref{tab:yields}}.
\begin{figure}[!b]
\centering
\includegraphics[width=0.55\textwidth]{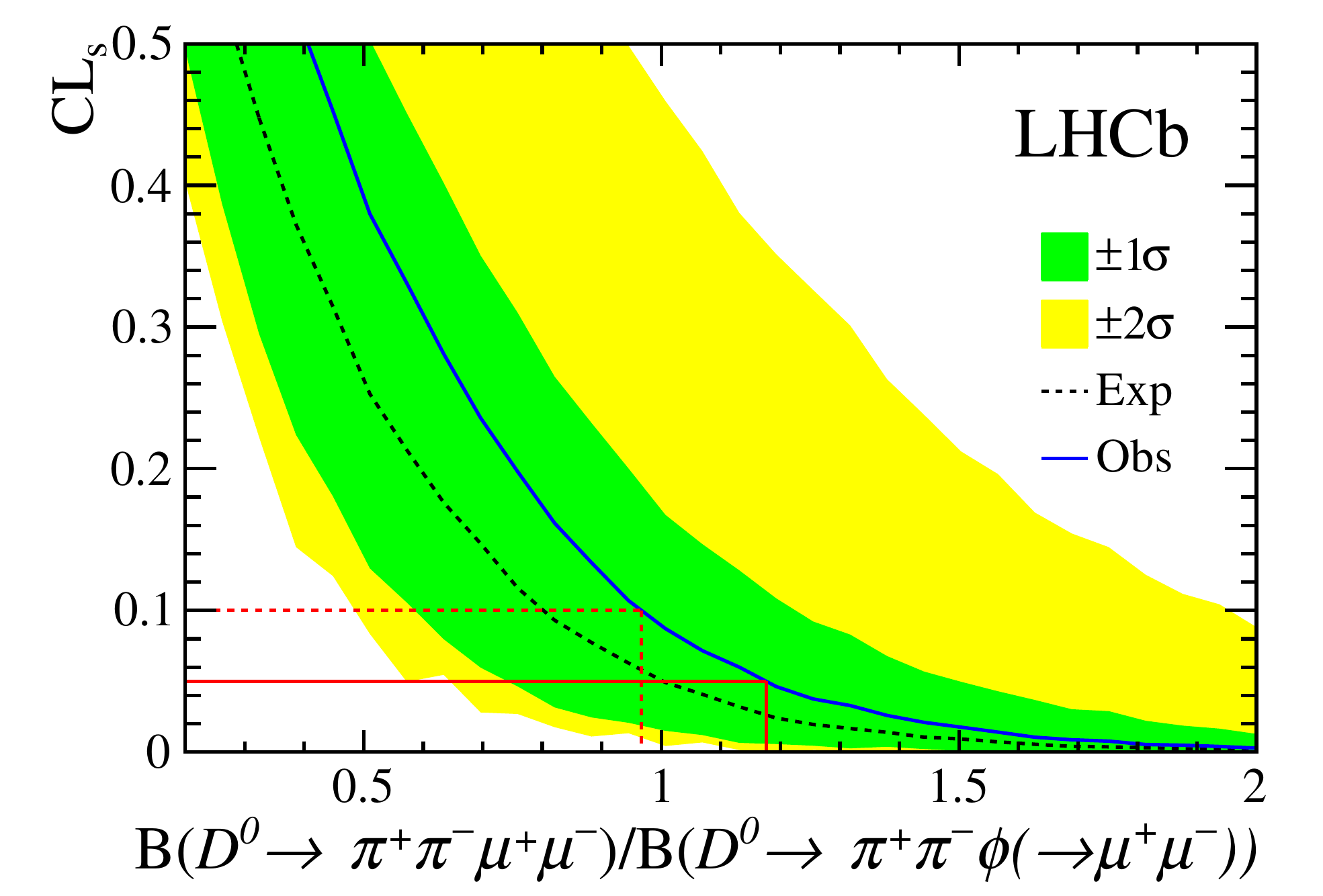}
\caption{\small{Observed (solid curve) and expected (dashed curve) \cls values as a function of $\BF(\Dppmm)/\BF(\Dppmmnorm)$. The green (yellow) shaded area contains 68.3\% and 95.5\% of the results of the analysis on experiments simulated with no signal. The upper limits at the 90(95)\% \cl are indicated by the dashed (solid) line.}}
\label{fig:cls}
\end{figure}
\begin{figure}[!h]
\centering
\includegraphics[width=0.55\textwidth]{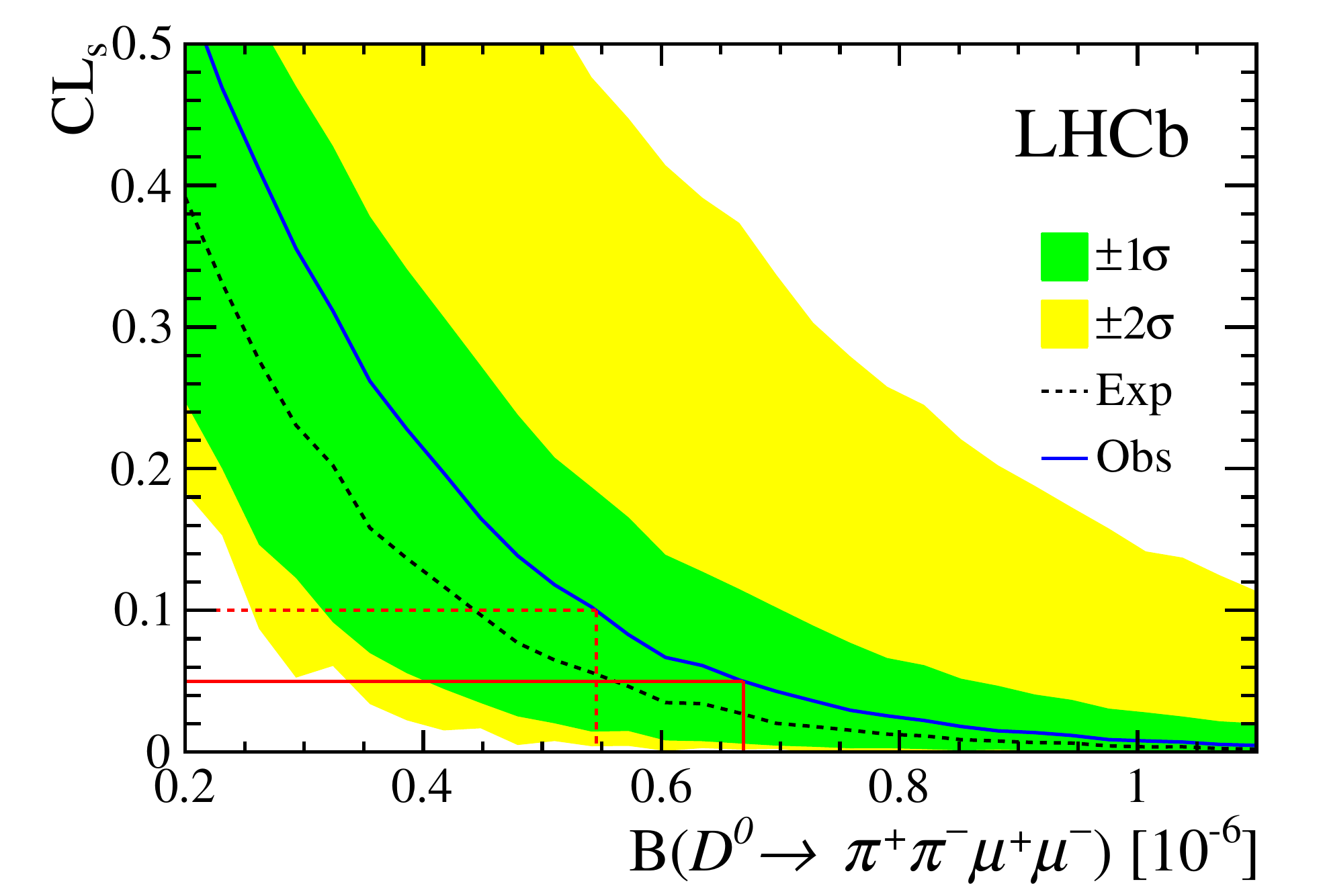}
\caption{\small{Observed (solid curve) and expected (dashed curve) \cls values as a function of \BF(\Dppmm). The green (yellow) shaded area contains 68.3\% and 95.5\% of the results of the analysis on experiments simulated with no signal. The upper limits at the 90(95)\% \cl are indicated by the dashed (solid) line.}}
\label{fig:clsabs}
\end{figure}
The observed distribution of \cls as a function of the total branching fraction ratio for \Dppmm is shown in Fig.~\ref{fig:cls}. A similar distribution for the absolute branching fraction is shown in Fig.~\ref{fig:clsabs}.
The upper limits on the branching fraction ratio and absolute branching fraction at 90\% and 95\% \cl and the p-values $(1-\clb)$ for the background-only hypothesis  are given in \tablename~\ref{tab:limits1} and in \tablename~\ref{tab:limits2}. The p-values are computed for the branching fraction value at which $\clsb$ equals $0.5$.
Despite the smaller event yield for \Dppmm relative to \Dppmmnorm, the upper limit on the total relative branching fraction is of order unity due to several factors. These are the low reconstruction and selection efficiency ratio in the signal region, the systematic and statistical uncertainties, and the extrapolation to the full $\Mmumu$ range according to a phase-space model.
\begin{table*}[!b]
\centering
\caption{Upper limits on $\mathcal{\BF}(\Dppmm)/\mathcal{\BF}(\Dz\to\pip\pim \phi (\to \mup\mu^-))$ at 90 and 95\% \cl, and p-values for the background-only hypothesis in each \Mmumu region and in the full \Mmumu range (assuming a phase-space model).}
\begin{tabular}{cccc}
\hline
 Region & $90\%$ & $95\%$ & p-value\\
\hline
 low-\Mmumu & 0.41 & 0.51 & 0.32\\
 high-\Mmumu & 0.17 & 0.21 & 0.12\\
 Total & 0.96 & 1.19 & 0.25\\
\hline
\end{tabular}
\label{tab:limits1}
\end{table*}
\begin{table*}[!t]
\centering
\caption{Upper limits on $\mathcal{\BF}(\Dppmm)$ at 90 and 95\% \cl in each \Mmumu region and in the full \Mmumu range (assuming a phase-space model).}
\begin{tabular}{ccc}
\hline
Region & $90\% \, [\times10^{-7}]$ & $95\% \, [\times10^{-7}]$\\
\hline
 low-\Mmumu & 2.3 & 2.9\\
 high-\Mmumu & 1.0 & 1.2\\
 Total & 5.5 & 6.7\\
\hline
\end{tabular}
\label{tab:limits2}
\end{table*}
It is noted that, while the results in individual \Mmumu regions naturally include possible contributions from \Drhoppmm since differences in the reconstruction and selection efficiency with respect to the four-body \Dppmm are negligible, the extrapolation to the full \Mmumu phase-space
depends on the four-body assumption. Distinguishing a $\rho$ component in the dipion mass spectrum requires an amplitude analysis which would be hardly informative given the small sample size and beyond the scope of this first search.

Contributions for non-resonant \Dppmm events in the normalisation mode \Mmumu window are neglected in the upper limit calculations. Assuming a branching fraction equal to the 90\% \cl upper limit set in the highest \Mmumu region, the relative contribution of the non-resonant mode is estimated to be less than 3\%, which is small compared with other uncertainties.
\afterpage{\clearpage}
\section{Conclusions}
\label{sec:conclusions}

A search for the \Dppmm decay is conducted using $pp$ collision data, corresponding to an integrated luminosity of 1.0 \invfb at $\sqrt{s}=7$ \tev recorded by the LHCb experiment. The numbers of events in the non-resonant \Mmumu regions are compatible with the background-only hypothesis. The limits set on branching fractions in two $m(\mumu)$ bins and on the total branching fraction, excluding the resonant contributions and assuming a phase-space model, are
\begin{eqnarray*}
\frac{\mathcal{\BF}(\Dppmm)}{\mathcal{\BF}(\Dz\to\pip\pim \phi (\to \mup\mu^-))} &<& 0.96 \, (1.19), \mathrm{\,\,\,at\,\,the\,\,90\,(95)\%\,\,\cl},\\
\mathcal{\BF}(\Dppmm) &<& 5.5 \, (6.7) \times 10^{-7}, \mathrm{\,\,\,at\,\,the\,\,90\,(95)\% \,\,\cl}.\\
\end{eqnarray*}
The upper limit on the absolute branching fraction is improved by a factor of $50$ with respect to the previous search~\cite{Aitala:2000kk}, yielding the most stringent result to date.
\clearpage
\section*{Acknowledgements}

\noindent We express our gratitude to our colleagues in the CERN
accelerator departments for the excellent performance of the LHC. We
thank the technical and administrative staff at the LHCb
institutes. We acknowledge support from CERN and from the national
agencies: CAPES, CNPq, FAPERJ and FINEP (Brazil); NSFC (China);
CNRS/IN2P3 and Region Auvergne (France); BMBF, DFG, HGF and MPG
(Germany); SFI (Ireland); INFN (Italy); FOM and NWO (The Netherlands);
SCSR (Poland); MEN/IFA (Romania); MinES, Rosatom, RFBR and NRC
``Kurchatov Institute'' (Russia); MinECo, XuntaGal and GENCAT (Spain);
SNSF and SER (Switzerland); NAS Ukraine (Ukraine); STFC (United
Kingdom); NSF (USA). We also acknowledge the support received from the
ERC under FP7. The Tier1 computing centres are supported by IN2P3
(France), KIT and BMBF (Germany), INFN (Italy), NWO and SURF (The
Netherlands), PIC (Spain), GridPP (United Kingdom). We are thankful
for the computing resources put at our disposal by Yandex LLC
(Russia), as well as to the communities behind the multiple open
source software packages that we depend on.

\addcontentsline{toc}{section}{References}
\ifx\mcitethebibliography\mciteundefinedmacro
\PackageError{LHCb.bst}{mciteplus.sty has not been loaded}
{This bibstyle requires the use of the mciteplus package.}\fi
\providecommand{\href}[2]{#2}

\end{document}